\documentclass{aa}
\usepackage{graphicx,natbib}
\bibpunct{(}{)}{;}{a}{}{,}
\begin{document}


\title{The ISOGAL field FC--01863+00035: Mid-IR interstellar extinction
and stellar populations \thanks{This is paper no. 15 in a refereed journal
based on data from the ISOGAL project}\fnmsep
\thanks{Based on observations with ISO, an ESA
project with instruments funded by ESA Members States (especially the
PI countries: France, Germany, the Netherlands and the United Kingdom)
and with the participation of ISAS and NASA; and on DENIS observations
collected at the European Southern Observatory, Chile} }

\author{B.W. Jiang\inst{1,2,3} \and A. Omont\inst{2} \and
S. Ganesh\inst{2,4} \and G. Simon\inst{5} \and F. Schuller\inst{2}}

\offprints{B.W. Jiang (bjiang@bnu.edu.cn)}

\institute{Department of Astronomy, Beijing Normal University, Beijing 100875, China \and Institut
d'Astrophysique de Paris, CNRS, 98 bis Bd Arago, F-75014 Paris, France
\and National Astronomical Observatories, Chinese Academy of
Sciences, Datun Rd.  No.20(A), Beijing 100012, China (on leave)
\and Physical Research Laboratory, Navarangpura, Ahmedabad 380009,
India \and GEPI, Observatoire de Paris, 21 Avenue
de l'Observatoire, F-75014 Paris, France}

\date{Received: July 18; accepted December 23, 2002}

\titlerunning{The ISOGAL field FC--01863+00035 }
\authorrunning{B.W. Jiang et al.}

\abstract{ A 0.35\degr $\times$ 0.29\degr field centered at $l$=--18.63\degr,
$b$=0.35\degr
 was observed during the ISOGAL survey by ISOCAM imaging at 7$\mu$m
and 15{\rm $\mu$m}.  648 objects were detected and their brightness
are measured.
By combining with the DENIS data in the near-infrared J and K$_{\rm S}$ bands,
one derives the extinction at 7{\rm $\mu$m} through ${\rm A_{K_{\rm S}}-A_7
= 0.35 (A_J-A_{K_{\rm S}})}$ which yields A$_{7}$/A$_{\it V}$ $\sim$0.03 from
the near-IR extinction values of van de Hulst--Glass (Glass 1999).  
The extinction
structure along the line of sight is then determined from the values
of J--K$_{\rm S}$ or K$_{\rm S}$--[7] of the ISOGAL sources identified as RGB or early
AGB stars with mild mass-loss. The distribution of A$_{\it V}$ ranges from
0 to $\sim$45 and it reflects the concentration of the extinction in
the spiral arms. Based on their locations in color-magnitude diagrams
and a few cross-identifications with IRAS and MSX sources, the nature
of objects is discussed in comparison with the case of a low
extinction field in Baade's Window. Most of the objects are either AGB
stars with moderate mass loss rate or luminous RGB stars. Some of them may be
AGB stars with high mass loss rate. In addition,
a few young stellar objects (YSOs) are present.
 }

\maketitle

\section{Introduction}

ISOGAL is an ISOCAM
survey at 7$\mu$m and 15{\rm $\mu$m}, at a spatial resolution in
pixel-field-of-view of 6\arcsec and sensitivity about 10mJy, of about
16 deg$^{2}$, toward the Galactic plane mostly interior to $|l| =
30\degr$ (Omont et al. 2002).  About 200 fields observed are well spread
in the inner bulge and in the Galactic disk. In combination with the
DENIS data (Epchtein et al. 1997), the
colors between 15$\mu$m, 7$\mu$m, K$_{\rm S}$, J, I in the ISOGAL-DENIS
catalogue (Schuller et al. 2002) allows a detailed study of cold stellar
populations. For example, this survey shall result in a practically
complete census of mass-losing AGB stars in the fields of the inner
bulge and in some parts of the Galactic disk. The stars at the RGB tip
may also be well characterized in the ISOGAL catalogue, as well as
nearby or massive young stellar objects~(YSOs). In addition to the
study of cold stellar populations, another goal of ISOGAL is to study
the Galactic structures in regions highly obscured through the inner
Galaxy with a sensitivity and spatial resolution about two orders of
magnitude better than IRAS.

FC--01863+00035 is one of the disk fields within the ISOGAL survey. In
order to avoid strong sources saturating the ISOCAM detectors, an
ISOGAL field is usually limited to a small $l \times b$ raster where
no bright IRAS objects exist. The field FC--01863+00035 covers an area of
about 0.1 deg$^{2}$ in the Galactic plane.  Unlike the fields studied
in the Galactic bulge by Glass et al. (1999) and Omont et al. (1999), this
disk field suffers serious interstellar extinction in the Galactic
plane.  We picked it to make a case study of ISOGAL data in the
Galactic disk, taking advantage of the recent availability of the
ISOGAL-DENIS PSC. This line of sight is interesting because it
crosses four spiral arms, with large values of visible extinction up
to 30 mag and beyond, corresponding to strong CO emission (Bronfman et al. 1989). Being much closer to the Galactic Center than the field of
the early study by P\'erault et al.(1996), at $l = -45\degr$, it is
more typical of the majority of ISOGAL fields. Being outside of the
tangential direction of the molecular ring, it avoids too strong
perturbations of the quality of ISOGAL data by star forming regions,
while keeping nevertheless a non negligible number of detected
YSOs. On the other hand, it is far enough from the Galactic Center so
that disk sources well prevail against bulge ones, although this
direction is neither very far from the far end of the bar structure
(e.g. Lopez-Corredoira et al. 2001). In addition to the discussion of
stellar populations detected by ISOGAL, which is more difficult than
in the bulge because of the larger uncertainty on the distance of the
sources, the main goal of this paper is to show how the combination of
ISOGAL and DENIS data allows to study the properties and the structure
of interstellar extinction in the inner Galactic disk.

\section{Observation and data reduction}

The observation by the 60-cm space telescope ISO took place towards
the center position of $l$=--18.63$\degr$ and $b$=+0.35$\degr$, i.e. ${\rm
\alpha=16^{h}50^{m}25.4^{s}~ and ~\delta=-43\degr58\arcmin29\arcsec
}$. The rasters covered a rectangular $l \times b$ area of $0.35 \times 0.29$
deg$^{2}$. The details of the ISOGAL observation procedure with
ISOCAM (Cesarsky et al. 1996) are described in Schuller et al. (2002).
The observations of the field FC--01863+00035 were performed
with 6$\arcsec$ pixels with the filters LW2 (5.0--8.5 $\mu$m,
$\lambda_{\rm ref}=6.75\mu$m) and LW3 (12--18$\mu$m, $\lambda_{\rm ref}=14.3\mu$m ). The log of ISOGAL and DENIS observations is given in Table 1.

\begin{table}[h]
\caption[]{Log of Observations. }
\begin{tabular}{lllll} \hline
         & Filter & Date & ISO ION & pfov \\ \hline
ISOGAL & LW2 &
1997.03.18 & 48801636 & 6\arcsec \\
ISOGAL & LW3 & 1996.09.26 & 31500236 & 6\arcsec \\
DENIS & I,J,K$_{\rm S}$ & 1996.03.25 & & 3\arcsec \\ \hline
\end{tabular}
\label{tablog}
\end{table}

\begin{figure*}[htb]
\centering \includegraphics[width=9.5cm]{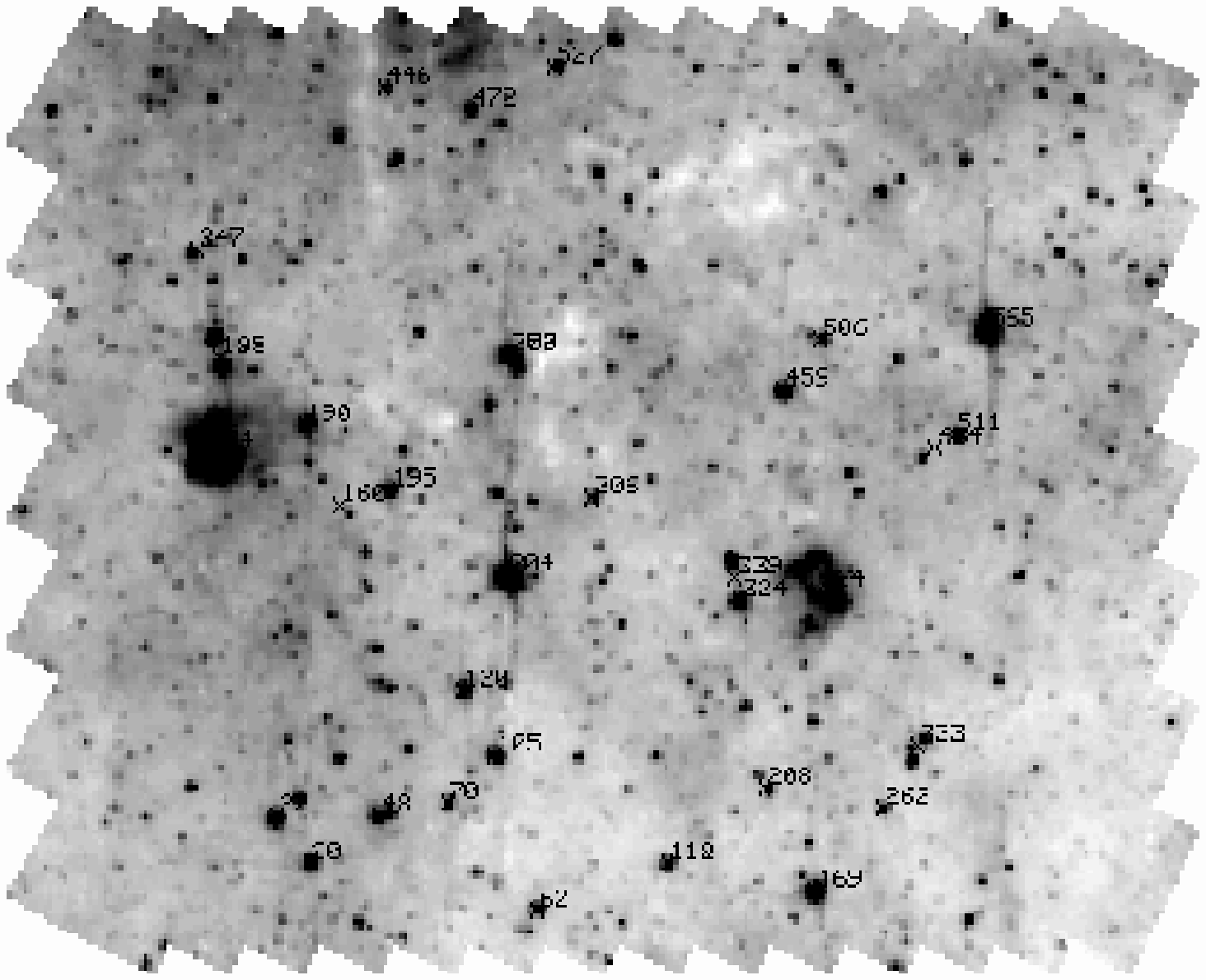} \nolinebreak
\hspace*{-0.5cm} \includegraphics[width=9.5cm]{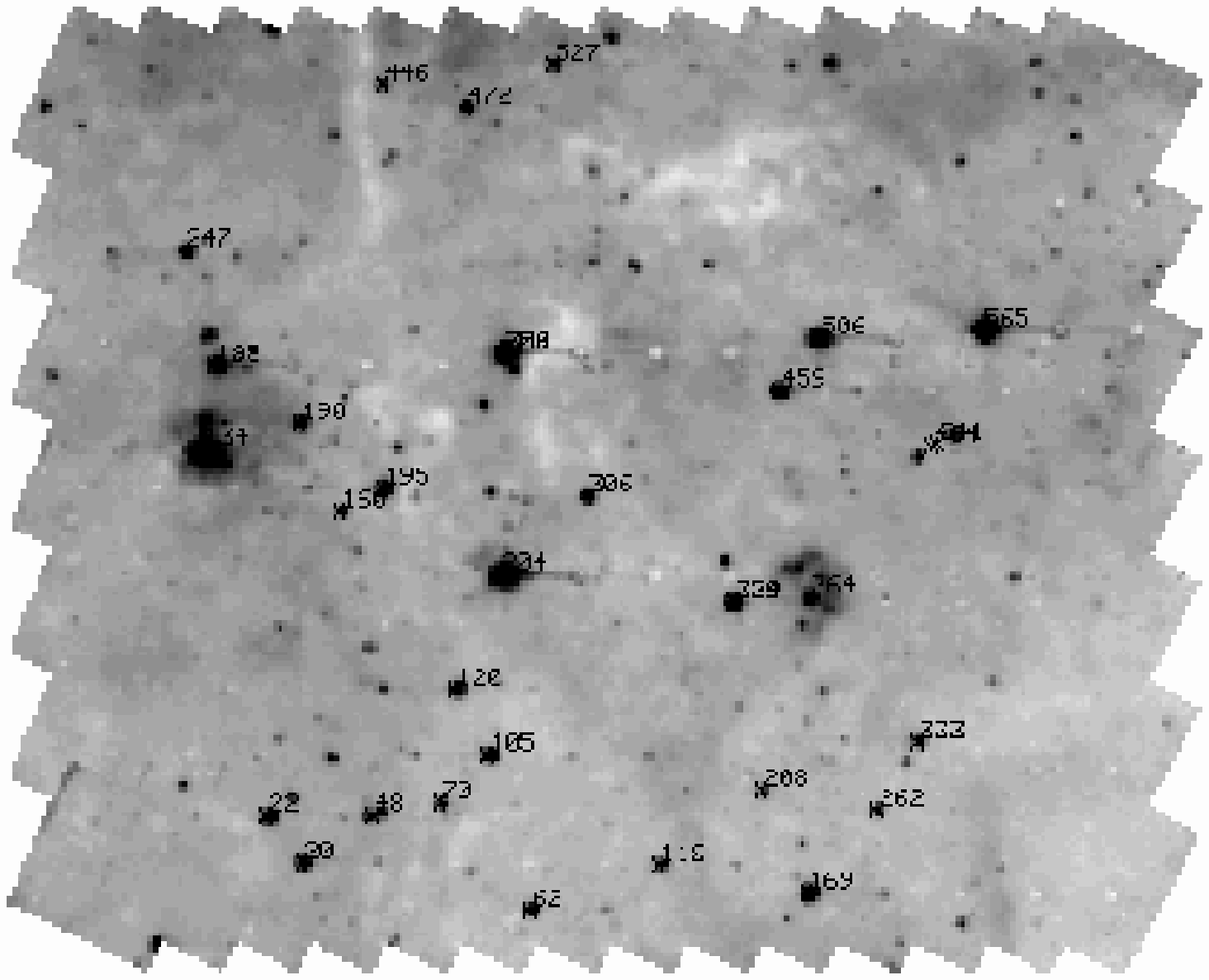}
\caption{The ISOCAM images of the field FC--01863+00035 in LW2 (left) and
LW3 (right) bands. Increasing $l$ is to the right and increasing $b$
is downwards. The sources with [7]$<$5.0 or [15]$<$5.0 are labeled
by the sequence numbers in the ISOGAL PSC catalog. (see Table \ref{tabiras} for those detected  by IRAS) }
\label{isolw}
\end{figure*}

The general processing of the ISOGAL data is described in detail in
 Schuller et al.(2002).  The reduction of the data from the OLP7.0
 (OffLine Processing) pipeline was performed by first using the CIA
 package (Ott et al. 1997). Dark currents are corrected and cosmic-rays
 are removed. Thereafter, a procedure is applied 
 to simulating the time behavior of the pixels of the
 ISOCAM detectors, eliminating the related artifacts and improving the photometry. The flat-field and image distortion are
 corrected. Conversion to magnitudes from
 the calculated flux density at 7$\mu$m and 15{\rm $\mu$m} is under the assumption
 that a Vega model, without circumstellar dust, corresponds to zero
 magnitude at respective wavelengths, i.e.  ${\rm
 [7]=12.38-2.5logF_{LW2}(mJy)}$ and ${\rm
 [15]=10.79-2.5logF_{LW3}(mJy)}$.  The current version of the images
 are shown in Fig.~\ref{isolw}. On average, the rms dispersion of the
 ISOGAL photometry (from repeated observations - see for example
 Schuller et al. 2002) is estimated to be generally less than 0.2
 magnitude with a small increase for the faintest sources that
 correspond to about the 50\% completeness limit.

\begin{figure*}[htb]
\begin{center}
\resizebox{15cm}{!}{\includegraphics[angle=90]{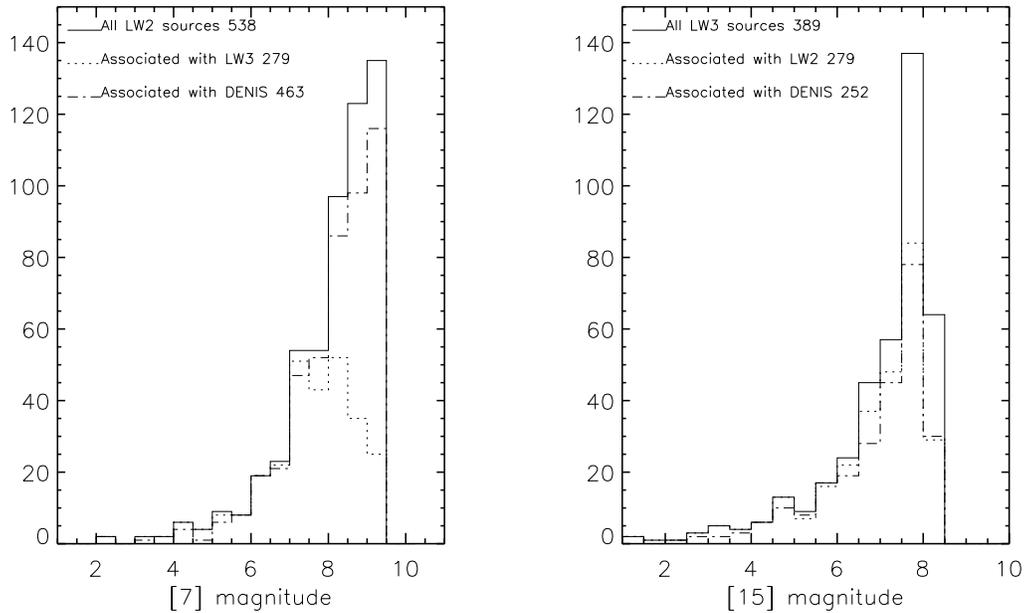}}
\caption{The sources detected in the ISOGAL observation. The solid
line histogram shows the distribution of all ISOGAL sources, the short
dash line shows the distribution of the sources associated with the other ISOGAL band and the
dash-dot line shows the distribution of the sources detected during the DENIS survey where only the
sources with good association flags are counted. The
left graph is the histogram of LW2 sources and the right one is of LW3
sources.  }
\label{hist715}
\end{center}
\end{figure*}

In the field FC--01863+00035, the numbers of point sources extracted are
538 and 389 respectively in LW2 and LW3 bands within the limits of
magnitude 9.38 in LW2 and 8.16 in LW3, which correspond to the flux
limits of about 15mJy and 11mJy, respectively (generally such limits
are chosen in the ISOGAL PSC such that they correspond to detection
completeness $\sim$50\% (Schuller et al. 2002) ).  There are in total 648
ISO sources, out of which 279 objects were detected in both LW2 and
LW3 bands, 259 objects detected in only LW2 band and 110 objects
detected in only LW3 band. Among the 279 LW2-LW3 associated sources,
257 sources have good association quality flags 3 or 4 and 21
sources have doubtful associations with quality flag 2. We will
discard the single association with quality flag 1 hereafter.  With the
association radius 5.4\arcsec, the number of LW2-LW3 spurious associations
should be $\sim$3. The ISO sources are distributed along brightness as
shown in Fig.~\ref{hist715}. As can be seen from Fig.~\ref{hist715}
the detection is certainly not complete to sources fainter than
magnitude 9.0 in LW2 or fainter than magnitude 7.5 in LW3. A general
tendency is that more brighter sources are detected in both LW2 and LW3
bands while more weaker sources are detected in only one band.

\begin{figure*}[htb]
\begin{center}
\resizebox{16cm}{!}{\includegraphics[angle=90]{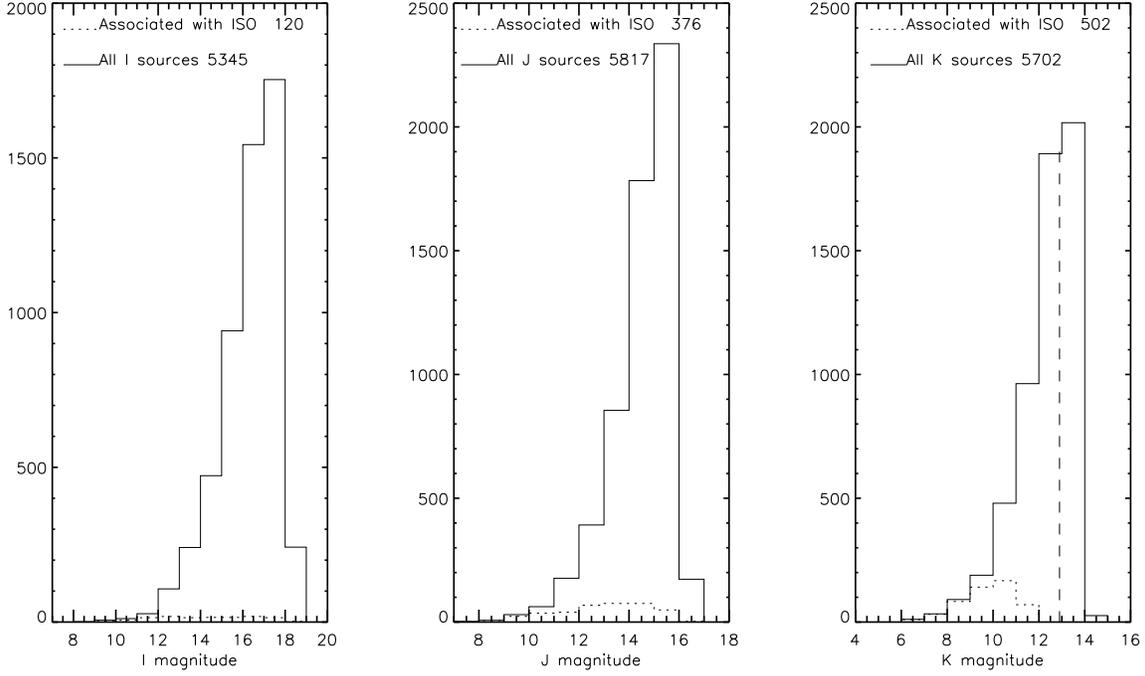}}
\caption{The DENIS sources in the ISOGAL field. All the sources 
detected by DENIS in the same field are distributed as
indicated by the solid line. Those associated with the ISOGAL objects
are represented by the short-dashed line and their number are also shown
in the upper part of the graphs. From left to right are
aligned bands I, J and K$_{\rm S}$. In the graph of the K$_{\rm S}$-band, a long-dashed vertical
line marks the limit K$_{\rm S}$=12.9mag used for searching DENIS association.}
\label{histijk}
\end{center}
\end{figure*}

In near-infrared, this ISOGAL field was also observed in the 2MASS survey in 
the J, H and K$_{\rm S}$ bands, but the data are not yet available. Contrarily, it was 
early observed  in a special observation of the
 DENIS survey in the I, J, K$_{\rm S}$ bands by a 1-m telescope at ESO, La Silla
 (Simon in preparation). The DENIS survey has much higher
 sensitivity (by typically 3 magnitudes in K$_{\rm S}$) as well as higher
 spatial resolution (by about 2$\sim$3 times) than the ISOGAL
 survey. The number of sources detected by DENIS is much larger and
 limited by confusion in the J and K$_{\rm S}$ bands. There are 5345, 5817 and 5702 sources
 in the I, J and K$_{\rm S}$ bands, respectively.  From the distribution of the
 brightness of these DENIS sources in this field as shown in
 Fig.~\ref{histijk}, the detections in the near IR bands are
 reasonably complete to I=16 mag, J=14 mag and K$_{\rm S}$=12 mag.

In order to keep the population of spurious DENIS-ISOGAL
cross-identifications below a few percents, the association was
limited to K$_{\rm S}$-detected sources with K$_{\rm S}$$<$12.9 (shown by a long-dash line
in the right panel of Fig.~\ref{histijk}), corresponding to a
density of K$_{\rm S}$ sources ${\rm n=36000/deg^2}$. Such a density limit is
systematically applied in the ISOGAL PSC for the DENIS association
radius. It is chosen such as ${\rm n\pi r_a^2=0.1}$, where ${\rm
r_a=3.5\arcsec}$ is the main association radius. The numbers of
sources associated with a DENIS source in the ISOGAL PSC are thus 479
objects in the LW2 band and 270 objects in the LW3 band. Among these
objects, 16 LW2 sources and 18 LW3 sources (in total 24 sources only) are
poorly associated with the DENIS objects, i.e. their association quality
factors are either 1 or 2 in the ISOGAL PSC (see for details from
Schuller et al. 2002) and they are dropped for later discussion. The
numbers of associations with quality flag equal to 3 are 9 in the LW2 band
and 10 in the LW3 band (15 sources in total); they mainly correspond
to association radius between ${\rm r_a}$ and 2${\rm r_a}$. The
associations with quality factors 4 or 5 (463 sources) are very
probably real associations with a proportion of spurious associations
less than $\sim$2\%. The associations with quality flag 3 still have
a good chance to be real, but with a larger proportion of spurious
associations; in this field however no such objects have association
radius larger than 5\arcsec.  Among the 538 LW2 objects, 463 (86\% of
all LW2 sources) are thus reasonably well associated with the DENIS K$_{\rm S}$
sources, 351 are also detected in the J-band and 109 in all DENIS
bands. Among the 389 LW3 objects, 252 (65\% of all LW3 sources) are
reasonably well associated with the DENIS K$_{\rm S}$ sources, 198 are also
detected in the J-band and 54 in all DENIS bands. Three DENIS sources
associated with ISOGAL have a bad quality in K$_{\rm S}$ because of saturation
(K$_{\rm S}$$<$6.5), and one J-associated source is saturated (J$<$8.0).  One
can check on the DENIS K$_{\rm S}$ and J images that five strong sources
present at the position of strong ISOGAL sources are missing in the
DENIS catalog because of saturation (see Sect. 4.3).  From
Fig.~\ref{histijk} it can be seen that most of the DENIS objects with
K$_{\rm S}$$<$9 with 11 exceptions were detected by ISO.  Among the ISOGAL sources
that are not associated with the DENIS sources, 31 are detected in
both LW2 and LW3 bands, 27 in only LW2 band and 87 in only LW3
band. There are 15 LW3 objects associated with K$_{\rm S}$ objects and
with association quality flag $\geq$ 3,  but are not
associated with LW2 sources. They may be spurious
cross-identification between DENIS and ISOGAL catalogues and are
dropped in the following discussions.

The details of the results including the astrometric, photometric and
association information of all the ISOGAL objects will be available on the
web in
the ISOGAL-DENIS PSC via http://vizier.u-strasbg.fr/viz-bin/VizieR and
http://www-isogal.iap.fr (Schuller et al. 2002).


\section{Interstellar extinction}

This field FC--01863+00035 at $b$=0.35$\degr$ is close to the Galactic
mid-plane where the interstellar extinction is serious. The measurement of
interstellar extinction becomes then impossible in optical at large
distances. Based on the optical observations of O-type to F-type
stars, the A$_{\it V}$ value at the direction of  $l$=--18$\degr$ and
$b$=--0.5$\degr$
which is close to the ISOGAL field, is estimated to be about 3
magnitude and approximately constant from a distance of $\sim$1\,kpc up
to $\sim$4\,kpc (Neckel \& Klare 1980). This may be a good reference for short
distances although the patchy distribution of interstellar matter in
the Galactic plane (Schultheis et al. 1999) may lead to some significant
difference from the extinction in the ISOGAL field
FC--01863+00035. However, the sources detected in the mid-infrared LW2 and LW3
bands extend to a much larger distance because of the much smaller
extinction at these wavelengths than in optical.

Our calculation of interstellar extinction is based on two
assumptions. One is that most of the objects in the ISOGAL field are
luminous RGB stars or AGB stars with moderate mass loss rate. These
objects are very bright with a luminosity of about one thousand to
several thousand solar luminosity, and cold with an effective
temperature lower than 4000\,K or so. In addition, stellar winds from
the photosphere have formed a circumstellar envelope for many of
them. So they are strong emitters in mid-infrared.  The
analysis of the ISOGAL fields at ($l$=0.0\degr~, $b$=1.0\degr)
 and in Baade's Window,
has actually found most of the sources there are RGB and AGB stars
whose colors are mildly reddened by interstellar extinction
(Omont et al. 1999; Glass et al. 1999). The other assumption is that the intrinsic
colors (J--K$_{\rm S}$)$_{0}$ of all these sources are approximately similar as well as
(K$_{\rm S}$--[7])$_{0}$, which is not true for the minority of objects with
large mass-loss.  According to the compilation of Wainscoat et al.(1992),
the value of (J--K)$_{0}$ of an M0III star is about 1.0 and of an M5
early AGB star is about 1.3 (see also van Loon et al. 2002 and references therein).  In one ISOGAL field Sgr I that is inside
the Baade's Window (Glass et al. 1999), the concentration of (J--K$_{\rm S}$)$_{0}$
and (K$_{\rm S}$--[7])$_{0}$ is very evident; for the objects with (K$_{\rm S}$--[7])$<$1.0,
the average (J--K$_{\rm S}$)$_{0}$ is
1.08 with standard deviation of 0.17~mag and the average (K$_{\rm S}$--[7])$_{0}$
being 0.20 with standard deviation of 0.23~mag (with the revised photometry of
the ISOGAL PSC, see Sect. 3.1) when the interstellar extinction
is subtracted as 0.2~mag in the K$_{\rm S}$-band.

\subsection{Extinction law at 7 and 15 micron}

The assumption to start to extract the extinction law in mid-infrared
is that the intrinsic color index (J--K$_{\rm S}$)$_{0}$ of most stars in the field is
the same so that their observed color J--K$_{\rm S}$ represents the interstellar
extinction.  Another presumption is the linear correlation between two
color indexes if they are both originated from interstellar
extinction, e.g. a linear correlation between the interstellar
reddening at J--K$_{\rm S}$ and K$_{\rm S}$--[7], and at J--K$_{\rm S}$ and K$_{\rm S}$--[15]. For the first
assumption, we had to remove some objects with apparently different
intrinsic color. In addition to RGB stars and AGB stars with moderate
mass loss, YSOs and AGB stars with high mass loss may be important
members of the mid-infrared objects. Their cold and thick
circumstellar envelopes absorb near-infrared radiation and radiate in
mid-infrared strongly and make them easily detected by the ISOGAL
survey. Their intrinsic color indexes (K$_{\rm S}$--[15])$_{0}$ and
(K$_{\rm S}$--[7])$_{0}$ [and even (J--K$_{\rm S}$)$_{0}$ for some of them] are redder
than RGB stars or early AGB
stars, and their color index [7]--[15] is also redder. Because the observed
values of J--K$_{\rm S}$, K$_{\rm S}$--[7] and K$_{\rm S}$--[15] are significantly influenced by
interstellar extinction, they can not be used as a criterion to pick
up the late AGB stars or YSOs.  On the other hand, the interstellar
extinction in the LW2 and LW3 bands is much smaller, the color index
[7]--[15] should be affected little by interstellar extinction. From
the ISOGAL results in the Baade's Window (Glass et al. 1999) and from
modeling (Ojha et al. 2002) we estimate that, for [7]--[15]$<$0.4,
the effect of circumstellar dust of AGB stars is negligible on
(K$_{\rm S}$--[7])$_{0}$, that it remains small ($\la$0.3) on (K$_{\rm S}$--[15])$_{0}$,
and that it is similarly negligible on (J--K$_{\rm S}$)$_{0}$ for
[7]--[15]$<$1.0. Therefore we exclude the stars with ${\rm [7]-[15]>
0.4 }$ or ${\rm [7]-[15]>1.0}$ in the discussion of the relation of
the colors K$_{\rm S}$--[7] and J--K$_{\rm S}$, respectively, with interstellar
extinction. Applying the same criteria for young stars also warrants
that the effect of circumstellar dust is negligible on their intrinsic
colors (see e.g. Bontemps et al. 2001).

If we set the intrinsic color (J--K$_{\rm S}$)$_{0}$ of most ISOGAL stars as
C$^{0}_{\rm JK_{\rm S}}$ and of (K$_{\rm S}$--[7])$_{0}$ as C$^{0}_{\rm K_{\rm S}7}$, and the
observed color indexes as C$_{\rm JK_{\rm S}}$=J--K$_{\rm S}$ and C$_{\rm K_{\rm S}7}$=K$_{\rm S}$--[7] 
accordingly,
the linear relationship is expected to be C$_{\rm K_{\rm S}7}$--C$^{0}_{\rm K_{\rm S}7}$
=\,$k$ (C$_{\rm JK_{\rm S}}$--C$^{0}_{\rm JK_{\rm S}})$. Based on model calculation
(e.g. Bertelli et al. 1994 ) and the observation of the field Sgr I, the
intrinsic color C$^{0}_{\rm JK_{\rm S}}$ of RGB stars or early AGB stars is
taken to be 1.2 on average, while the intrinsic color index
C$^{0}_{\rm K_{\rm S}7}$ of these stars is not well known (however, see van Loon et al. 2002). By leaving
C$^{0}_{\rm K_{\rm S}7}$ as a variable and assuming that the observed C$_{\rm
K_{\rm S}7}$ of sources with $[7]-[15] < 0.4$ is mainly caused by
interstellar extinction, a robust linear fitting method, which
minimizes absolute deviation and is insensitive to large departures
for a small number of points, was used to fit the data and the result
is displayed in the color-color diagram (Fig.~\ref{ccdjkk7}, left).
The resulted values for this linear fitting to the points decoded by
pluses and triangles in Fig.~\ref{ccdjkk7} are: $k$=0.35 and
C$^{0}_{\rm K_{\rm S}7}$=0.34 (assuming C$^{0}_{\rm JK_{\rm S}}$=1.2 ).

 The processing of K$_{\rm S}$--[15] is similar to that of K$_{\rm S}$--[7].  The results
are then {\rm $k=0.39$ and C$^{0}_{\rm K_{\rm S}15}$=0.39
(assuming C$^{0}_{\rm JK_{\rm S}}$=1.2) (Fig.~\ref{ccdjkk7}, right). In
Fig.~\ref{ccdjkk7}, the sources with [7]--[15]$>$0.4 are displayed by
diamonds though they were abandoned during the fitting process to both
the K$_{\rm S}$--[7] and K$_{\rm S}$--[15] with J--K$_{\rm S}$.  Their large deviation to the right
side from the fit line further proves that they are intrinsically
redder. However, there could be a residual effect of circumstellar
dust on the value of K$_{\rm S}$--[15] even for sources with [7]--[15]$<$0.4,
especially for sources with large values of J--K$_{\rm S}$ which are more
luminous; it could lead to an overestimation of $k$ by maybe
$\sim$10-20\%.

 For a blackbody radiation at temperature 4000K typical of an M0 RGB
star, the Planck function yields C$^{0}_{\rm K_{\rm S}7}$=0.35 and
C$^{0}_{\rm K_{\rm S}15}$=0.44 when the flux density for K$_{\rm S}$=0.0 is taken as
 6.65$\times10^{24}$ W/m$^{2}$/Hz (Schuller et al. 2002) ,
 which is consistent with the fitting results C$^{0}_{\rm K_{\rm S}7}$=0.34
and C$^{0}_{\rm K_{\rm S}15}$=0.39.  Our values of C$^{0}_{\rm K_{\rm S}7}$ and C$^{0}_{\rm K_{\rm S}15}$
are in reasonable agreement with those from Glass et al. (1999) in Baade's
Windows
when one takes into account the correction of $-$0.45 magnitude to be
applied to the preliminary ISOGAL photometry used by Glass et al. (1999)
(Schuller et al. 2002).

\begin{figure*}[htp]
\centering
\includegraphics[width=9.5cm]{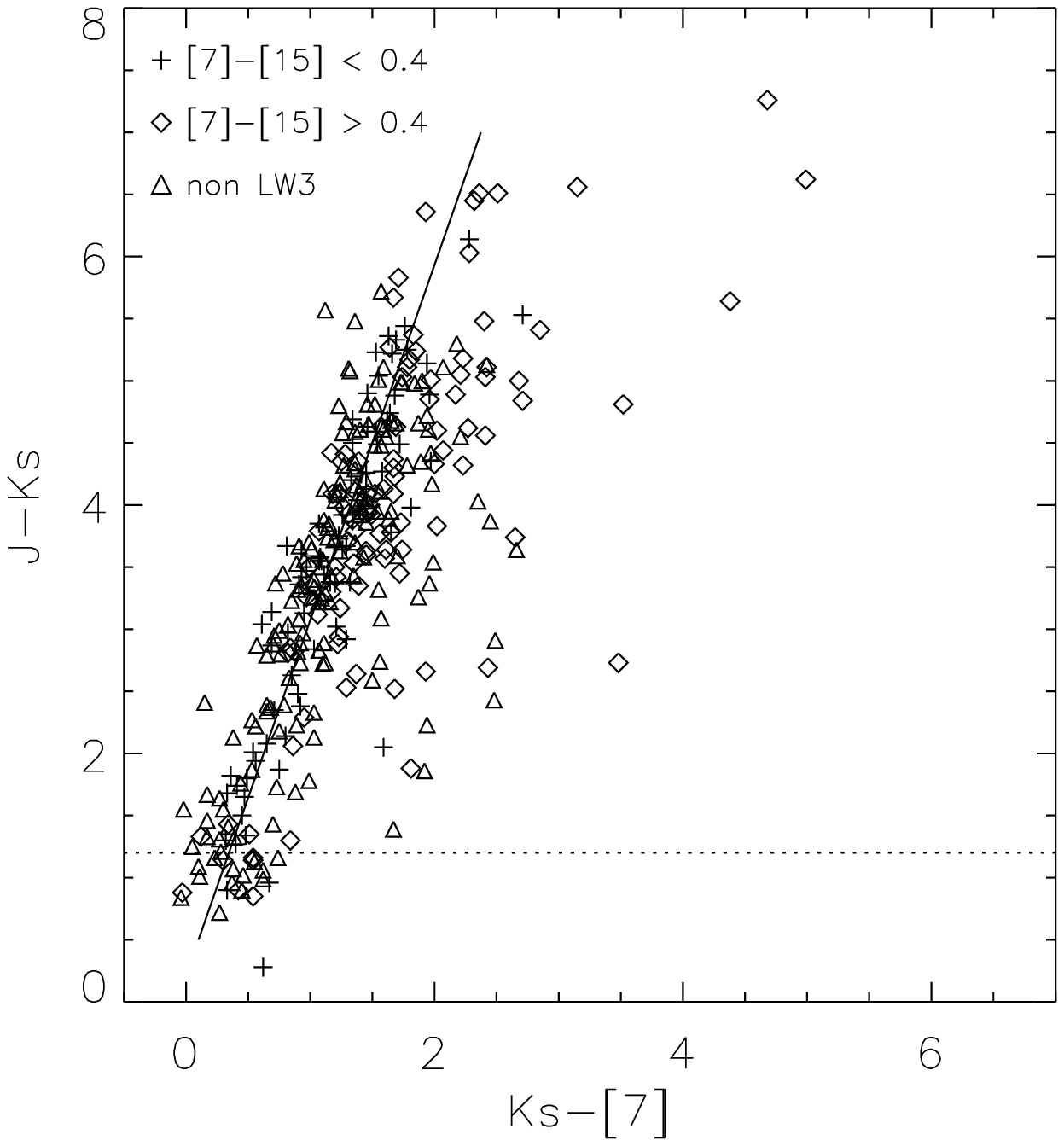} \nolinebreak \hspace*{-1cm}
\includegraphics[width=9.5cm]{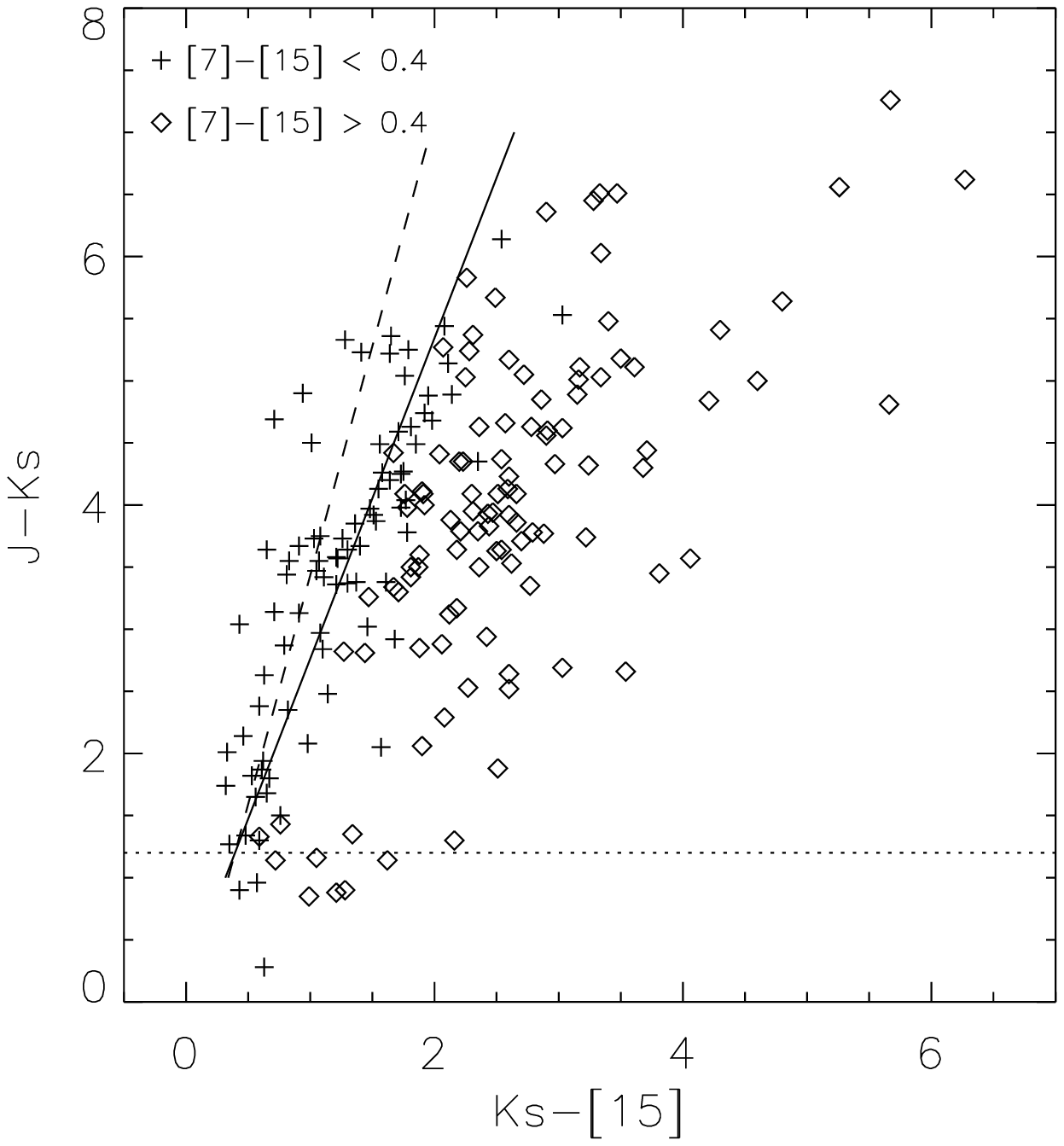}
\caption{Color-color diagram of J--K$_{\rm S}$ vs. K$_{\rm S}$--[7] and J--K$_{\rm S}$
vs. K$_{\rm S}$--[15]. The sources detected in the LW3 band are divided into two
groups, one with ${\rm [7]-[15] < 0.4}$ decoded by pluses which were
used for linear fitting and the other with ${\rm [7]-[15] > 0.4}$
decoded by diamonds which were not used for linear fitting. In the
left graph, the sources which are not detected in the LW3 band but are
used in linear fitting are decoded by triangles. The robust linear
fitting result is shown by solid line.  The horizontal dot line labels where
J--K$_{\rm S}$=1.2 which is assumed intrinsic color J--K$_{\rm S}$ for the sources used in 
fitting. In the right graph, a dash line represents the
result of combining the 7 to 15$\mu$m opacity
 ratio from Hennebelle et al. (2001) with
the extinction values of J, K$_{\rm S}$ and 7~$\mu$m in Col. 2 of
Table.~\ref{exttab} (i.e. A$_{\rm 15}$=0.044).  }
\label{ccdjkk7}
\end{figure*}

The linear coefficients between C$_{\rm JK_{\rm S}}$ and C$_{\rm K_{\rm S}7}$ or
C$_{\rm K_{\rm S}15}$ define the extinction values at 7 and 15 microns which
depend on the extinction values in near infrared. With the extinction
values in near infrared given by van de Hulst(1946), Glass(1999) and Rieke \& Lebofsky(1989), the corresponding values at 7 and 15 microns are
calculated and listed in Table \ref{exttab} by adopting the
coefficients $k=0.35$ and $k=0.39$ between C$_{\rm JK_{\rm S}}$ and C$_{\rm
K_{\rm S}7}$, and C$_{\rm JK_{\rm S}}$ and C$_{\rm K_{\rm S}15}$, respectively.

\begin{table}[h]
\caption[]{Extinction values in the LW2~(7{\rm $\mu$m}) and LW3~(15{\rm
$\mu$m}) bands}
\begin{tabular}{lccc} \hline
Band & \multicolumn{3}{c}{$A_{\lambda}/A_{v}$} \\ &
 vdH\footnotemark[1]\ & ISG\footnotemark[2] & R \& L\footnotemark[3] \\
 \hline J & 0.245 & 0.256 & 0.281 \\
      K$_{\rm S}$ & 0.087 & 0.089 & 0.112 \\
      7 & 0.032 & 0.031 & 0.053\footnotemark[4] \\
     15 & 0.026 & 0.025 & 0.046\\ \hline
\end{tabular}

\vspace*{1mm}
\begin{small}
\noindent \footnotemark[1]Values for J and K$_{\rm S}$ derived from
van de Hulst (1946) curve\\
\footnotemark[2] Values for J and K$_{\rm S}$ derived from
Glass (1999) (update of van de Hulst's values)\\
\footnotemark[3] Values for J and K$_{\rm S}$ derived
by Rieke \& Lebofsky (1989) for stars towards the Galactic center\\
\footnotemark[4] 7 and 15$\mu$m extinction values
are then derived from the slopes $k$ of the straight lines of
Fig.~\ref{ccdjkk7} (see text): A$_{\rm K_{\rm S}}$--A$_7$=0.35(A$_{\rm J}$-A$_{\rm K_{\rm S}}$),\,
 A$_{\rm K_{\rm S}}$--A$_{15}$=0.39(A$_{\rm J}$--A$_{\rm K_{\rm S}}$).
\end{small}
\label{exttab}
\end{table}

While the extinction values in the near-infrared from Rieke \&
Lebofsky (1989) are higher, the ones preferred by Glass (1999) are
just an update of those of van de Hulst(1946) and thus the two latter
sets of values are quite close to each other. Similarly, the
extinction values at 7 and 15$\mu$m derived from Rieke \& Lebofsky
{\rm $(A_{7} \approx A_{15} \approx 0.05)$ } are higher than those
inferred from Glass--van de Hulst, i.e. A$_{7}\approx0.03$A$_V$ and
$A_{15}\approx0.025$A$_V$ (see Table \ref{exttab}). As generally in
other ISOGAL papers, we prefer Glass' values for the reasons exposed
in Glass (1999), except possibly for the fields close to the
Galactic Center and in star forming regions. To compare the values we
have derived for the mid-infrared extinction with previous ones, we
will distinguish the case of 7~$\mu$m where the fitting of
Fig.~\ref{ccdjkk7} (left) is very robust, from the one of 15~$\mu$m
which is more uncertain. In both cases one should note that the
averages of the extinction on the wavelength range of the LW2 and LW3
broad bands yield slightly larger values than at 7 and 15{\rm $\mu$m},
respectively, which are both close to a minimum of the extinction
curve.

At 7~$\mu$m the most widely used reference for the extinction value is probably Mathis (1990) which is similar to Draine \& Lee (1984). 
They used similar extinction law in near infrared to Rieke \& Lebofsky (1989),
 and estimated the infrared extinction from an extrapolation of the optical
extinction law and near-infrared observational data, yielding A$_7$
$\sim$0.020A$_{\it V}$. This value is slightly lower than the one we derived, 0.03A$_{\it V}$. Such a difference may arise
from the uncertainty in the ISOGAL photometry or in the fitting
procedure in Fig.\ref{ccdjkk7}. It is probably included as well within the
uncertainty of the extrapolation by Mathis (1990).  We will use our value, 0.03A$_{\it V}$, in the following of this paper, keeping in view its uncertainty. Let however note that the higher value we could derive from the near-infrared extinction law of Rieke \& Lebofsky (1989) (Table 2), is also consistent with the results derived by Lutz (1999) (A$_{7} \approx$0.045)
who used the hydrogen recombination lines and the same near-infrared
extinction law towards the Galactic Center.

The extinction ratio A$_7$/A$_{15}$ is given as $\sim$1.3 by Mathis (1990). However, the value of Draine \& Lee (1984) is almost twice smaller. The best value for this ratio has probably been derived by Hennebelle et al. (2001) from an analysis of infrared dark clouds from the ISOGAL survey.  They give 0.7$\pm$0.1 for the clouds away from the Galactic Center, which is similar to Draine \& Lee (1984).
The value we derived for this ratio in Table 2, Col. 2, 1.2, is more compatible with Mathis (1990) than with Hennebelle et al. (2001). 
In Fig.\ref{ccdjkk7} (right),
the 15 $\mu$m value deduced from the Hennebelle et al. ratio is shown by a dashed line
derived from the extinction values at J, K$_{\rm S}$ and 7$\mu$m taken from
Col. 2
of Table~\ref{exttab}. It is evident that their result implies more
extinction at 15$\mu$m. However, let us stress that our direct 
fitting is somewhat
uncertain at 15 $\mu$m because of the smaller number of sources and of the possibility of residual effects
of circumstellar reddening on K$_{\rm S}$--[15], especially for the distant
luminous sources with large extinction. Therefore, we consider that
the 15 $\mu$m value deduced from Hennebelle et al. (2001) (dashed line in
Fig.\ref{ccdjkk7} (right)) is still compatible with the data of
our fitting in Fig.\ref{ccdjkk7} (right, full line), and that our
data cannot provide a really accurate value of the extinction at 15
$\mu$m.
Nevertheless, we will use the values
derived from our fitting
in the subsequent sections, since such differences on the small
extinction at 15 $\mu$m are practically negligible in the following
discussions. One has to keep in mind that the extinction law may vary
with the directions due to the inhomogeneous distribution of
the interstellar matter in the Galactic plane.

\subsection{Extinction structure along the line of sight}

By assuming the intrinsic color index C$^{0}_{\rm JK_{\rm S}}$ is the same
for all the objects which were detected in both J and K$_{\rm S}$ bands and
with [7]--[15]$<$1.0 if detected in both LW2 and LW3 bands as well, the
interstellar extinction to individual objects can then be calculated
from the observed C$_{\rm JK_{\rm S}}$.  Theoretical calculation shows that
C$^{0}_{\rm JK_{\rm S}}$ doesn't differ much for late-type RGB stars and
early AGB stars which are the major components of the ISOGAL
sources. Though there may be some foreground main-sequence stars
and early-type AGB stars
with intrinsic bluer color than late-type RGB stars that would bring about the
underestimation of ${\rm A_{\it V}}$, the number of such sources should be
small and they should be mostly at small distance to be detectable by
ISOGAL as seen in Fig.~\ref{ccdjkk7} for those with J--K$_{\rm S}$$<$1.0.
While many of the ISOGAL sources are detected in the K$_{\rm S}$-band and not
detected in the J-band, the value of ${\rm A_{\it V}}$ may then be inferred from
C$_{\rm K_{S}7}$ according to the extinction law derived above. This
method is right only if the non-detection in the J-band is caused by large
interstellar extinction. Because some of the non-detections in the J-band
may come from serious absorption by circumstellar dust of AGB
stars or YSOs, the estimation from C$_{\rm K_{S}7}$ may overestimate the
real interstellar extinction, in particular when the index C$_{\rm
K_{\rm S}7}$ is large, e.g. larger than 2 that results in $A_{V}\sim30$. So
when inferring the extinction value from K$_{\rm S}$--[7], we have excluded the
sources with [7]--[15]$>$0.4 because of the risk they present the
occurrence of large mass loss. For the same reason, we have excluded
the sources with [7]--[15]$>$1.0 when inferring the extinction value
from J--K$_{\rm S}$.  However, we have thus altogether discarded a large proportion of
sources with the largest extinction.

We adopted the value of intrinsic color index C$^{0}_{\rm JK_{S}}$ as 1.2
and the extinction values listed in Col. 2 of
Table~\ref{exttab}. The global distribution of ${\rm A_{\it V}}$ is shown in
Fig.~\ref{histav}d, where values of ${\rm A_{\it V}}$ from J--K$_{\rm S}$ or K$_{\rm S}$--[7]
are distinguished by the dash or dot lines respectively, while the
summation of the two types is represented by the solid line.  It can
be seen, as expected, that the sources not detected in the J-band
experience higher extinction than those detected in the J-band on
average. All the sources not detected in the J-band have ${\rm A_{\it V} >
10}$ and peak at about 28.  In Fig.~\ref{histav}, we have added to the bin A$_{\it V}$=0--2
all the sources with J--K$_{\rm S}$$<$1.2. Such sources should not be
AGB or RGB-tip stars, but nearby earlier stars, mostly K giants with
a few A-B stars. Most of them must have  A$_{\rm V}<$2--3. We have checked 
that the corresponding values of I--J are consistent with J--K for all these
19 sources, except for one where there is a problem with the DENIS 
I--J associations. As expected, all of them are associated with a 
visible GSC2--2 star. 
The combined visible/near-infrared colors are consistent with 
K-early or M giants 
for most of them, with, however, an identified A2 star.

\begin{figure*}[htbp] \resizebox{\hsize}{!}{\includegraphics{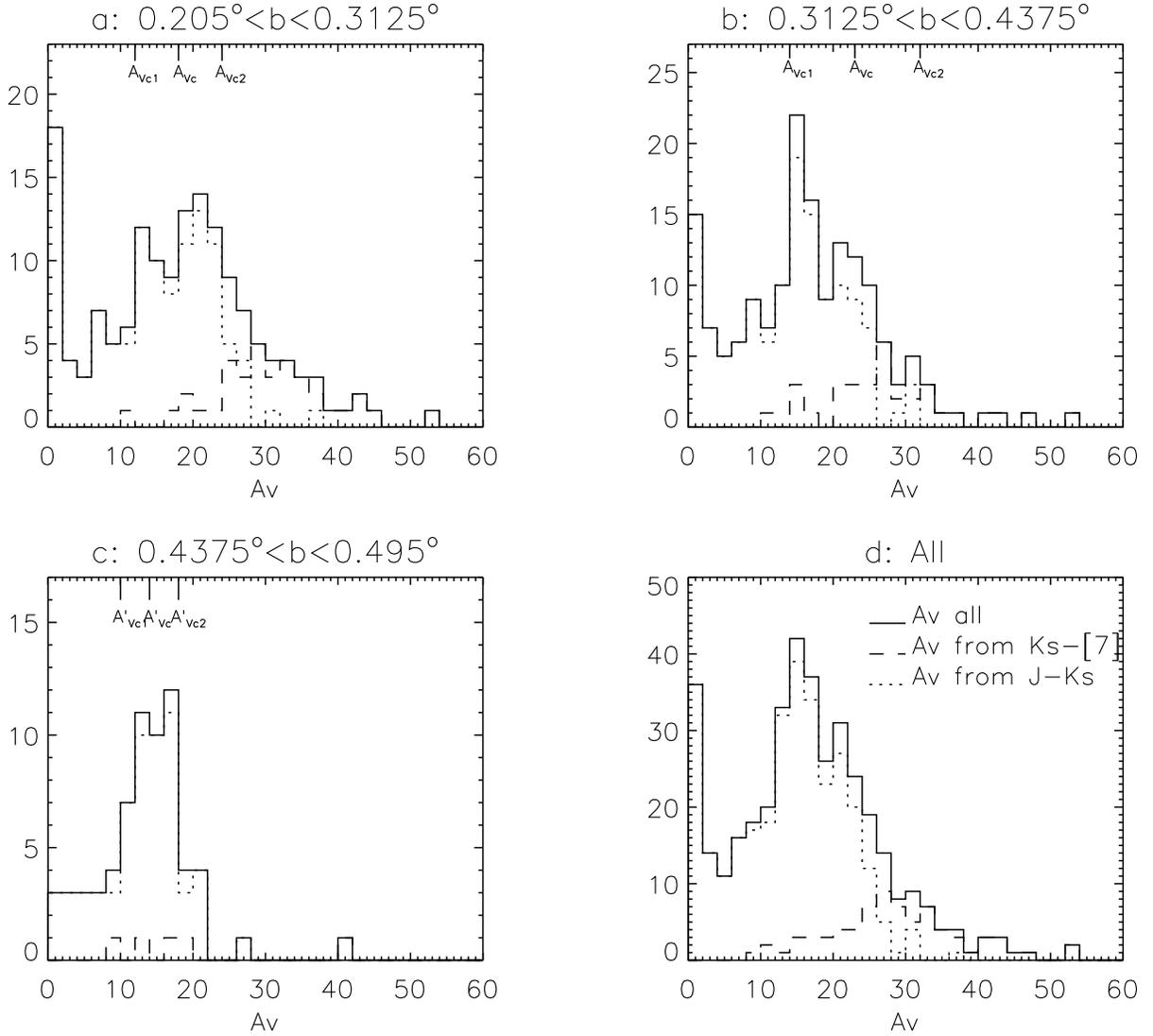}}
\caption{Histograms of the interstellar extinction ${\rm A_{\it V}}$,
which are derived either from J--K$_{\rm S}$ (dot line) or from K$_{\rm S}$--[7] (dash
line) when no data in the J-band are available (sources with [7]--[15] $>$
0.4 are excluded), and from the addition of these two types (solid
line). Figures a, b and c display the sources with $b$ in the quoted ranges
corresponding to the overlap of the ISOGAL field with the pointings of the
CO survey data by Dame et al.(2001). Figure d shows the extinction 
distribution of the sources spanning the whole $b$ range in this field. 
 A$_{\rm Vc}$ represents the average value of the
accumulated extinction expected from the interstellar gas just beyond the 
Centaurus arm (last line of Table 3); A$_{\rm Vc1}$ represents the starting 
 value of the extinction by the Centaurus arm as deduced from 
the sharp rise of 
the ${\rm A_{\it V}}$ histogram; and A$_{\rm Vc2}$ is the symmetric 
value of A$_{\rm Vc1}$  
with respect to A$_{\rm Vc}$, which could correspond to the 
ending value of the 
extinction by the arm. 
The same method cannot be applied to subfield ``c'' because the ISOGAL 
range is about half, much smaller
than the total range of the CO pointing; A$_{\rm Vc'}$ is instead defined 
as the average 
value of the peak of the histogram.   } 
\label{histav} 
\end{figure*}

\begin{table*}[htb] \caption{Average interstellar extinction A$_{\it V}$ in
different distance and $b$ ranges (inferred from Dame at al. 2001 
and Bloemen et al. 1990) }
\begin{tabular}{c|ccc|ccc|ccc|ccc}\hline 
& \multicolumn{3}{c|}{Nearby} & \multicolumn{3}{c|}{``Centaurus''} & \multicolumn{3}{c|}{``Norma''} &\multicolumn{3}{c}{``Tangent point'' } \\ 
Distance(kpc) & \multicolumn{3}{c|}{0--3} & \multicolumn{3}{c|}{3--6 or 7--11} & \multicolumn{3}{c|}{6--8} & \multicolumn{3}{c}{7--10} \\ 
V$_{\rm LSR}$(km/s) & \multicolumn{3}{c|}{$>$0} & \multicolumn{3}{c|}{$[0,-60]$} & \multicolumn{3}{c|}{$[-60,-100]$} & \multicolumn{3}{c}{$[-110,-150]$} \\ \hline
$b$\footnotemark[1] & a & b & c & a & b & c & a & b & c & a & b & c \\   
A$_{\it V}$(H$_{2}$ via CO) & 1.3 & 0.6 & 0.6 & 16.6 & 13.5 & 8.5  & 5.2 & 6.6 & 3.1 &  5.6 & 3.3 & 1.4 \\ 
 A$_{\it V}$(HI)/A$_{\rm CO}$ &\multicolumn{3}{c|}{1} & \multicolumn{3}{c|}{0.25} &\multicolumn{3}{c|}{0.60} &\multicolumn{3}{c}{0.70} \\ \hline
\multicolumn{13}{c}{Accumulated extinction} \\ \hline
Distance(kpc) & \multicolumn{3}{c|}{0--3} & \multicolumn{3}{c|}{0--6} & \multicolumn{3}{c|}{0--7\footnotemark[2] } & \multicolumn{3}{c}{0--10} \\
A$_{\it V}$(H$_{2}$+HI) & 2.5 & 1.2 & 1.2 & 23.3 & 18.1 & 11.8  &  31.7 & 28.8 & 16.8  & 41.2 & 34.3 & 19.0 \\\hline 
\end{tabular} 

\footnotemark[1]{The range of $b$ of the three CO subsamples are: 
a: 0.1875\degr$<b<$0.3125\degr, b :0.3125\degr$<b<$0.4375\degr, 
c: 0.4375\degr$<b<$0.5625\degr. }
\footnotemark[2]{Assuming that the CO emission is due to the near side of the Norma arm; otherwise, the 6--7 kpc extinction should be much smaller}
\label{tabav} \end{table*}

In addition to the concentration at A$_{\rm V}\sim$0, the sources are
distributed along ${\rm A_{\it V}}$ unevenly. After a dip at ${\rm
A_{V}}$$\approx$4--6, there is first a progressive increase of their
number up to ${\rm A_{\it V}}$$\approx$12, and then a steeper rise until a
relatively sharp maximum at ${\rm A_{\it V}}$$\sim$14--16. After that,
their number decreases rather regularly up to ${\rm A_{\it V}}$$\sim$50.

We think that the uneven distribution of ${\rm A_{\it V}}$ reflects the
 inhomogeneities in the distribution of the interstellar medium,
 partly along the line of sight with the crossing of the molecular
 ring and of several arms, but probably mainly perpendicular to the 
line of sight 
 across the observed ISOGAL field. In order to see the influence of
 the spiral arms, we estimated the extinction values to the arms from
 the distribution of the emission in the radio lines of CO and HI. The
 kinematical distance to the interstellar clouds can be inferred from
 their radial velocity. The line of sight in the
direction $l$=-18.63\degr~ and $b$=0.35\degr~ touches the outer edges of 
the Sagittarius,
the Centaurus and the Norma arm at respectively 1, 4, and 6 kpc, runs
through the bulge and then reaches the far side of the Norma arm at
about 13 kpc.
Indeed, most of the line of sight, between 3 to 13 kpc, is
 in the molecular ring, including the tangential point at $\sim$8kpc
 (v$\sim$ -130km/s).  

As for the whole Galactic disk, there exist $^{12}$CO data for our
field taken from the Milky Way survey at the coarse resolution of
0.125\degr (7.5\arcmin) (Dame et al. 2001). Our 0.35\degr$\times$0.29\degr~ field
thus implies nine pointings of this survey at $l=18.625\degr\pm0.125$\degr
and $b=0.375\degr\pm0.125$\degr. A rapid look at these CO emission data
shows that it corresponds mainly to radial velocities characteristic
of the molecular rings and more precisely, of the Centaurus and Norma
arms. However, there is an ambiguity for the emission between -70 -- -80
km/s which could be attributed either to the near side or to the far
side of Norma. Especially in the Centaurus arm, the CO emission
displays a strong gradient with respect to $b$ across this field
(Table \ref{tabav}). In order to take into account this gradient in the
discussion of the distribution of A$_{\it V}$, we will consider
separately the three $b$ ranges of the pointings of the CO survey
(Table \ref{tabav}). For each $b$ range of the pointings we computed
the CO integrated intensity, W$_{\rm CO}$, averaged over the three $l$
pointings, for each velocity interval roughly corresponding to the
different spiral arms (Table \ref{tabav}). Then we estimated the
corresponding A$_{V}$ shown in Table \ref{tabav}, adding a contribution
from the HI regions from the HI survey of Bloemen et al.(1990).

We estimated ${\rm A_{\it V}}$ shown in Table~\ref{tabav} by adopting the
conversion factor $ {\rm 1.8 \times 10^{20} cm^{-2} (K~km~s^{-1})^{-1}
}$ from the CO integrated line intensity W$_{\rm CO}$ 
to H$_{2}$ column density (Dame et al. 2001) and
the factor $ {\rm 10^{21} molecules~ cm^{-2}~ mag^{-1} }$ from H$_{2}$
column density to ${\rm A_{\it V}}$. This estimation, in particular of the
HI column density, suffers some uncertainty from integrating the
velocity on unclear contours of the paper that could be about 30\%. 
The W$_{\rm CO}$ to N$_{\rm H2}$ conversion factor is also known 
to be rather uncertain (Bachiller \& Cernicharo 1986; Harjunpaa \& Mattila
1996).

Similarly, we split our sample of ISOGAL
sources in three unequal parts corresponding to
these $b$ ranges (Table \ref{tabav}). For each subsample, we have represented in Fig.~\ref{histav} the histogram of the
distribution of A$_{V}$. We note, as expected, important differences
in the total ranges of these three distributions reflecting the
gradient of the extinction with $b$.

In view of discussing the correspondence between the A$_{\it V}$
distribution from the ISOGAL sources and the determination from the
interstellar gas, let us stress the difficulty that any spatial
inhomogeneity in the CO intensity smaller than the large CO beam is
smoothed with the present CO data. Therefore, it is impossible to
estimate the actual spatial dispersion of A$_{\it V}$ within a spiral
arm. It is certainly significantly larger than the dispersion,
$\sim$20\%, between the three $l$ pointings for the same $b$ range. In
each of the three histograms of Fig.~\ref{histav} the average value of
the accumulated extinction expected from the interstellar gas just
beyond the Centaurus arm is, A$_{\rm Vc}$=23, 18 and 12, respectively
(Table \ref{tabav}). Looking at
the values of the sharp rise of the A$_{\it V}$ distribution
due the Centaurus arm, with A$_{\rm Vc1}$=12, 14 and 10, respectively, 
one sees that
these are 20-50\% lower than A$_{\rm Vc}$. A natural interpretation
is that this difference essentially represents the dispersion of the
extinction through the Centaurus arm. 
The case of the subfield with 
$b>$0.4375\degr~ is special because the $b$ sample observed by ISOGAL, 
0.4375\degr$<b<$0.495\degr~ is about half the total range of the 
CO pointing, 0.4375\degr$<b<$0.5625\degr. Because of the strong 
gradient with $b$, it is likely that the average extinction in the 
region observed by ISOGAL is larger than the one in the total CO 
beam which yields A$_{\rm Vc}$=12 (see Table \ref{tabav}). Indeed 
the main feature of Fig.~5c is approximately symmetric with respect 
to A$_{\rm Vc'}$=14, which should be close to the actual value of 
the average of A$_{\it V}$ in the region observed by ISOGAL. 
 However, it is unclear in this case whether the main
contribution of Norma is absent because it comes from the far side
with very few sources behind it, or it is included in the pedestal
between A$_{\it V}$=18 and A$_{\rm V}=$22, together with the extinction of
a few sources behind the tangential point region.

The situation is less clear for the other two subsamples with
smaller $b$ and larger extinction. It is true that the distribution
again extends beyond the value A$_{\rm Vc2}$, symmetric of A$_{\rm Vc1}$,
 with respect to A$_{\rm
Vc}$, as expected from such a dispersion of the A$_{\it V}$
distribution through the Centaurus clouds and the additional extinction beyond Centaurus. But there is a large decline
of the number of the sources towards large A$_{\it V}$, making the
distribution very asymmetric with respect to A$_{\rm Vc}$.  The loss of
sources may be explained by the very large value of A$_{\it V}$ which
prevents the detection at 7$\mu$m or even in the K$_{\rm S}$-band 
of many sources.

Finally, we note that for the three subsamples, the total range of
values of A$_{\it V}$ inferred from the ISOGAL sources is consistent
with the maximum accumulated extinction along the line of sight
expected from the interstellar gas, including the dispersion of A$_{\rm
V}$ values within clouds, the Norma ambiguity and the loss of sources with very large A$_{\it V}$.

Despite the many uncertainties in the relation between W$_{\rm CO}$
and the ISOGAL colours and in the interpretation of the A$_{\it V}$
distribution, one feature remains intriguing. Before the sharp rise at
A$_{\rm V}\sim$10--14 that we associate with the Centaurus arm, there
is a distribution of a smaller number of sources from A$_{\rm V}\sim$6
 to $\sim$12. The local and Sagittarius arm emission of CO is quite
unable to account for such extinctions. The only explanation is that
such sources are located within the Centaurus arm, behind the first
dust layers. The subsequent very sharp rise suggests a narrow
distribution of the bulk of the Centaurus extinction. The appreciable
number of sources with A$_{\rm V}\sim$6 to $\sim$12 suggests that
the first Centaurus layers of dust are distributed over a rather large
distance with a substantial source density. This implies a large
width for the Centaurus ``arm'', which is consistent with the broad CO
velocity distribution, and/or an over-density of ISOGAL sources inside
it, which could be mostly young/massive K giants.

\section{Stellar populations}
\subsection{Sources detected in the J-band}
For the ISOGAL sources which are detected in both J and K$_{\rm S}$ bands
during the DENIS observation a correction for interstellar extinction
can be applied to observed color indexes such as K$_{\rm S}$--[7], K$_{\rm S}$--[15] and
[7]--[15].  The group of extinction values we used are in Col. 2
 of Table~\ref{exttab}. The color magnitude diagrams
(K$_{\rm S}$--[7])$_{0}$/[7]$_{0}$ and (K$_{\rm S}$--[15])$_{0}$/[15]$_{0}$, are shown in
Figs.~\ref{cmdk77} and \ref{cmdk1515} for the sources detected in both J
and K$_{\rm S}$ bands.  They are compared with the diagrams of the ISOGAL sources in
the Baade's Window Sgr I where the interstellar extinction is much
smaller. Specifically, the extinction in the K$_{\rm S}$-band in the Sgr I field is
taken to be 0.2 magnitude and no extinction is taken into account
there for the LW2 and LW3 bands (Glass et al. 1999).  In the
color-magnitude diagram [7]--[15]/[15] of our field FC--01863+00035  (see 
Fig. \ref{cmd71515}), we
have not applied the correction for interstellar extinction because it
is relatively small in both [7]--[15] and [15] (especially in [7]--[15]), 
and not certain for both.

\begin{figure}[htbp]
\resizebox{\hsize}{!}{\includegraphics{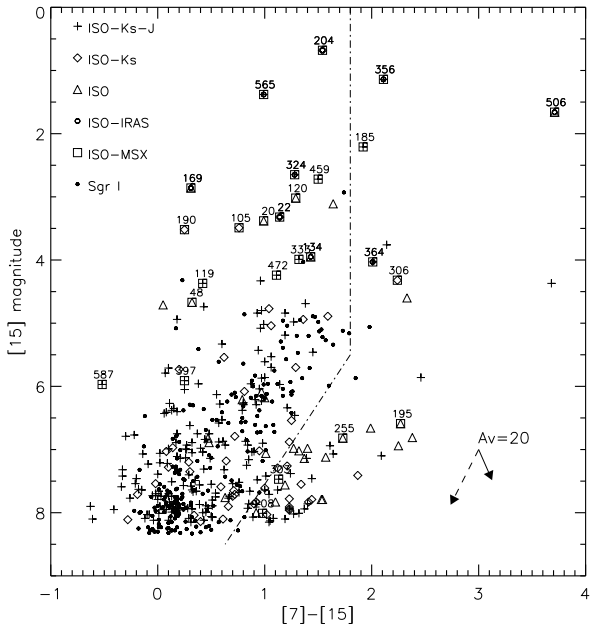}}
\caption{Color-magnitude diagram [7]--[15] vs. [15]without dereddening. The sources in the
field FC--01863+00035 are denoted by crosses for associations with both J
and K$_{\rm S}$ bands, by diamonds for associations only with the K$_{\rm S}$-band and
triangles for non-associations with DENIS. Those identified with the
IRAS PSC objects are circled, and with MSX point source catalogue are
squared.   The
objects identified with IRAS and/or MSX catalogues are numbered at the
top by their sequence in the ISOGAL field catalogue.
({\bf The sources in the SgrI field are decoded by dots.}) Extinction arrow
is displayed at the lower-right corner for A$_{\it V}$=20 mag, with
the two determinations of A$_{15}$ of Fig.\ref{ccdjkk7} (right). Dot-dashed
line represent the approximate borders between YSOs and
late-type stars, suggested by
Felli et al. (2000). }
\label{cmd71515}
\end{figure}

\begin{figure*}[htbp]
\includegraphics[angle=90,width=18cm]{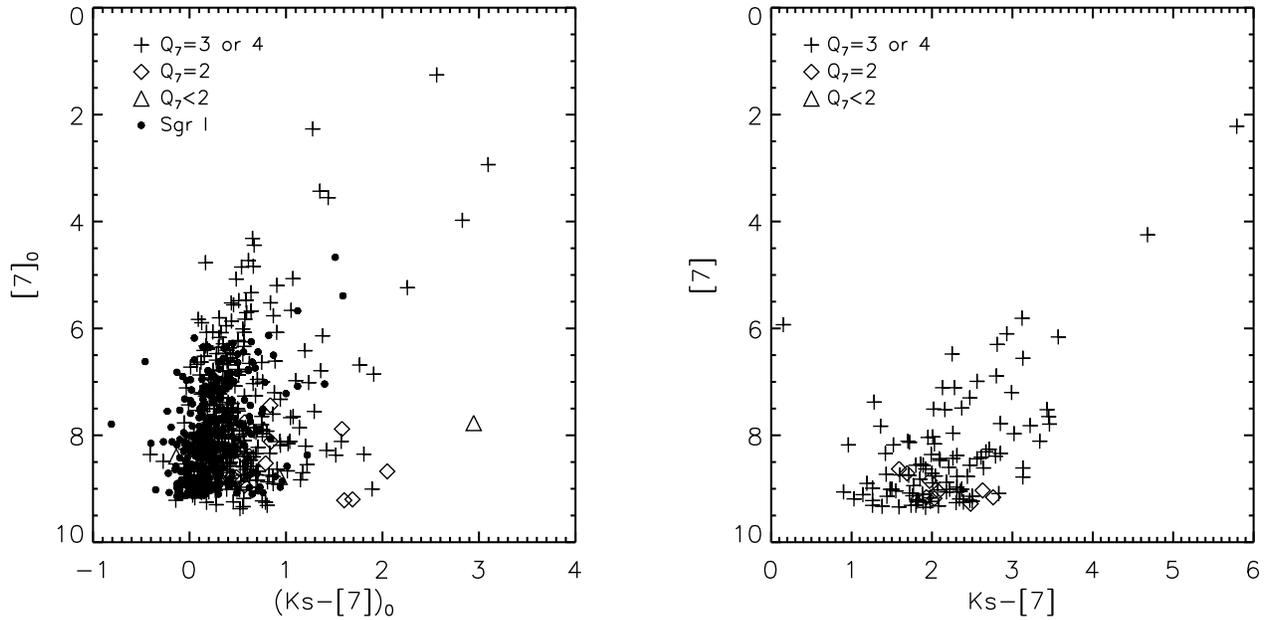}
\caption{Color-magnitude diagram K$_{\rm S}$--[7] vs. [7]. 
Left panel: 
Sources detected in both J and K$_{\rm S}$ bands and thus corrected for
interstellar extinction. The objects with good, moderate and poor 
quality flags at 7$\mu$m
are decoded by pluses, diamonds and triangles respectively. 
 The Sgr I objects are
again represented by dots and corrected for the extinction by 0.2mag
in the K$_{\rm S}$-band. 
Right panel: Sources not detected in the J-band and thus not
corrected for interstellar extinction. The pluses, diamonds and triangles
decode the sources with good, moderate and poor quality flags at 7$\mu$m. }
\label{cmdk77}
\end{figure*}

In these color-magnitude diagrams, the distribution of most of the
ISOGAL objects of the field FC--01863+00035 detected in the J-band is very similar
to that of Sgr I, with some small differences.  The cluster of stars
at the tip of RGB which is seen in the Sgr I field around [7]--[15] =
${\rm C_{715}=0.2, [15]=8.0}$ or (K$_{\rm S}$--[15])$_{0} = {\rm
C^{0}_{K_{\rm S}15}=0.3, [15]_{0}=8.0}$, does not appear clearly among the
LW3 sources in the disk field FC--01863+00035. Instead there is a loose
branch of stars along the line ${\rm C_{715}=0.2}$ or ${\rm
C^{0}_{K_{S}15}=0.3}$ with width of about $\pm$0.2 magnitude which are
probably also RGB stars.  Because the sources in the disk field
FC--01863+00035 span a wider range of distance than the concentration
mostly in the bulge in the case of the Baade's Window, their apparent
luminosity in bands LW2 and LW3 scatters more.  The intermediate AGB
stars, which are fainter by one or two magnitudes in absolute K$_{\rm S}$ or
M$_{\rm bol}$ and lose mass at lower rate than very luminous AGB stars
(Omont et al. 1999), appear in the region next to RGB branch with a
redder color ${\rm C^{0}_{715} }$. The stripe from ${\rm C_{715}=0.5,
[15]=7.5}$ to ${\rm C_{715}=1.0, [15]=6.0}$ coincides well with the
location of intermediate AGB stars in Sgr I. Upper along this sequence
are the luminous AGB stars with redder color, approximately in the
same region as Mira variables in the Sgr I field
(Glass et al. 1999). Several stars are brighter than magnitude 4 at [15],
which is lacking in the Sgr I field. These stars are mostly identified
in the IRAS PSC catalogue due to their strong flux density at 12 {\rm
$\mu$m}. They are discussed in a separate subsection. Many of them are
clearly much closer than the bulge distance and illustrate the fact
that the AGB sequence is not as well defined as in bulge fields
because of the wider spread in distances.

\subsection{Sources not detected in the J-band}

Non-detection in the J-band by DENIS can be due to three reasons. The most
frequent one is that interstellar extinction absorbs most of the
stellar radiation in the J-band. The second is that the J absorption is
due to a very thick and cold circumstellar envelope. Finally a few
very strong J sources may not figure in the DENIS catalogue because
they are badly saturated (Sect. 2).  The reddest color indexes J--K$_{\rm S}$
of objects detected by ISOGAL and DENIS J and K$_{\rm S}$
 in this field go up to about 7,
corresponding to A$_{\it V}$=35 mag if (J--K$_{\rm S}$)$_{0}<$1.5. The J
detections are relatively complete only to J--K$_{\rm S}$ = 5, i.e. A$_{\it V}$
$\sim$25, which approximately corresponds to the cumulated average
extinction of the Centaurus and Norma arms (see Sect. 3.2). The non 
detections in the J-band
are mainly due to a very large extinction which may arise either
because the star is behind a concentration of interstellar matter in
the Centaurus (or Norma) arm, or because it lies much farther in the
molecular ring. Their observed color K$_{\rm S}$--[7] is on average
about one magnitude redder than those detected in the J-band. This
tendency can be seen from the color-magnitude diagram of Fig.~\ref{cmdk77}. 
If the rough relationship derived from the objects
detected in the J-band between K$_{\rm S}$--[7] and J--K$_{\rm S}$ is applicable to the
non-detections in the J-band, then one magnitude difference in K$_{\rm S}$--[7]
corresponds to 3 magnitude difference in J--K$_{\rm S}$, i.e. about 18 magnitude
difference in A$_{\it V}$.

There are some J-nondetections whose location in the color-magnitude
diagram can not be explained by interstellar extinction alone. They
are bright in the LW2 or LW3 band, e.g. [7]$<$7.0 and [15]$<$6.0.  Their
color index [7]--[15] is around 1.0 that indicates much circumstellar
matter. They are good candidates for luminous Mira variable. These
stars are at the center of Fig.~\ref{cmd71515}, right-upper of
Fig.~\ref{cmdk77}.  

Several J-nondetections are not so bright as the
candidates for Miras. They are on the contrary a little faint in the LW3
band, from about [15]=7.0 to [15]=8.5. Their color index [7]--[15] is
usually redder than 1.0. These could be YSOs judged from the
relatively low luminosity and excess emission in the LW3 band. In
Fig.~\ref{cmd71515} they are in the middle-lower corner and a similar
position in Fig.~\ref{cmdk1515}.

\begin{figure}[htbp]
\resizebox{\hsize}{!}{\includegraphics{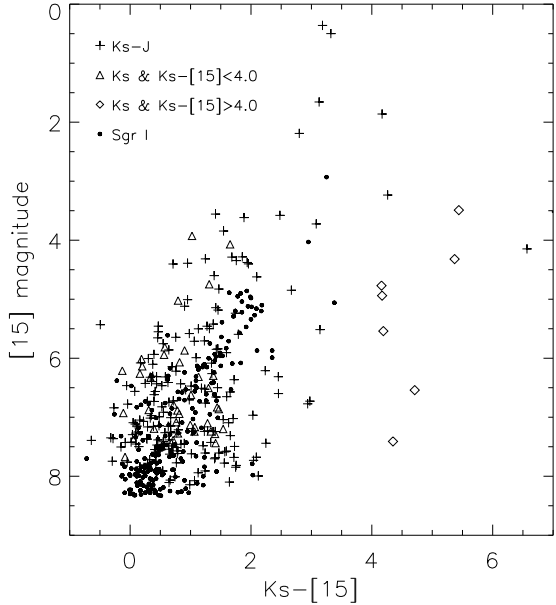}}
\caption{Color-magnitude diagram K$_{\rm S}$--[15] vs. [15]. 
 The sources detected in both J
and K$_{\rm S}$ bands (crosses) are corrected for interstellar extinction
according to their J--K$_{\rm S}$ values. Those not detected in the J-band with K$_{\rm S}$--[15]$<$4.0
(triangles)
are corrected for interstellar extinction according to their K$_{\rm S}$--[7] values, 
and
their K$_{\rm S}$--[15] and [15] are corrected for interstellar extinction.
The objects that were not
detected in the J-band and K$_{\rm S}$--[15]$>$4.0 (diamonds) {\bf are not corrected for
interstellar extinction} and their observed colors and magnitudes are
shown. }
\label{cmdk1515}
\end{figure}

\subsection{Sources not associated with DENIS}

The sources which are not associated with DENIS can be divided into
three groups according to the ISO bands of detection: in both LW2 and LW3
bands, only LW2 band and only LW3 band.

The sources detected in both LW2 and LW3 bands, but not in DENIS, are
denoted in Fig.  \ref{cmd71515} by triangles. There are five objects,
no. 20, 48, 120, 169 and 198 which are so bright to cause saturation
in the
DENIS observation and rejection from the DENIS catalogue. From their locations in Fig.~\ref{cmd71515},
sources no. 48, 169 and 198 may be nearby RGB star or an intermediate
AGB star with relatively thin circumstellar envelope; sources no. 20
and 120 are redder and brighter, thus may be luminous AGB stars with
thick circumstellar envelope. 

The non-detection of the other sources is due
to their faintness in the DENIS bands. The faint 15$\mu$m ISO sources with [15]
$>$ 6.0 not detected in DENIS are partly mixed with those detected in
the K$_{\rm S}$-band and not detected in the J-band in Fig.~\ref{cmd71515} that were
suspected to be candidates for YSOs, again due to their red color and
low luminosity.  Their non-association with DENIS can be accounted for
by heavy absorption and coldness of circumstellar disk of YSOs, plus
possibly interstellar extinction. However, their nature needs confirmation
since most of the reddest and weakest sources in Fig.~\ref{cmd71515}
([7]-[15]$>$1.5) have at least one ISOGAL magnitude of poor quality (as
well as the sources with (K$_{\rm S}$-[7])$_{0} \geq$2.0 and [7]$_{0}\leq$7.5
in Fig.~\ref{cmdk77}).

The sources detected only in the LW2 band are mostly faint with [7] $<$
8.0 (see Fig.\ref{hist715}). If they are not artifacts, the non-detection in LW3 can be
understood by the lower sensitivity of ISOCAM  in the LW3 band than in the LW2 band,
the incompleteness in the LW3 band to faint objects, and
because the spectral energy distribution (SED) of the objects do not
rise steeply from 7{\rm $\mu$m} to 15{\rm
$\mu$m}.  As the DENIS survey is almost complete to K$_{\rm S}$=11.0 mag, their
color index K$_{\rm S}$--[7] would be greater than 2.0 and possibly very large,
i.e. A$_{\rm V}>$ 30 since the limiting [7] magnitude is taken to
be 9.38. They must be distant, for example, further than the near side of
the Norma arm. On the other hand, the objects were detected in LW2 at
such large distance, which means they must be bright in LW2. The
study of ISOGAL variables in the Galactic bulge showed that the AGB
variables are distinguishable from other M-type giants by their high
7{\rm $\mu$m} luminosities (Schultheis et al 2000; Schultheis \& Glass 2001). So they are 
possibly AGB variables.

The sources detected only in LW3 are mostly fainter than magnitude
7.0. The incompleteness of the ISOGAL survey in LW2 to faint objects
can account for part of the non-detection in LW2.
However, the reality of weak unassociated LW3 sources should be 
confirmed. There may be some real sources that are candidates for YSOs. The low
temperature of YSOs peaks their radiation in middle and even far
infrared. The non-detection in shorter wavelengths can be understood by
their weak emission in these bands. If they were RGB or AGB stars, the
higher sensitivity of ISOCAM in LW2 should bring them to the detection
limit in LW2. More observation of these sources would reveal their
identity.

\subsection{Sources identified in IRAS or MSX PSC}

Although the IRAS photometric results of these faint sources are
mostly uncertain at 60 and 100{\rm $\mu$m} (Table 4), the difference
in the SED in the IRAS four bands between AGB stars and YSOs is
generally clear. The YSOs rise more steeply in SED from 12 to 100
micron than AGB stars. According to the IRAS colors, there are three
YSO candidates \#134, \#364 and possibly \#506 ( or IRAS16464-4359,
IRAS16469-4348 and IRAS16472-4351) which have steep SED in the IRAS
bands.  A further support comes from the ISOGAL imaging, i.e.
around the first two YSO candidates extended nebulae are clearly
visible in the ISOCAM images in both LW2 and LW3 bands (see
also Fig.~\ref{isolw}).  In addition,
sources \#364 and \#506 lie in a region of the [7]--[15]/[15] diagram
assigned to YSOs by Felli et al.(2000), and \#134 lies slightly left of
the borderline between YSOs and late-type stars and it is still
acceptable to be a YSO. However, for \#506, the MSX color D--E appears
anomalously blue for a YSO. 

All the other IRAS objects, except \#169,
are probably AGB stars with high mass loss rate. The first criterion
is that their IRAS colors mimic such stars.  Moreover, the IRAS
variability indexes of \#204 and \#565 are 83 and 79 respectively, 
which seems to
point out to variable late-type stars. One object
\#22~(IRAS~16459-4354) is an OH maser star (Sevenster et al. 1997).  All
are in the region where the large-amplitude long-period variables are
found in Fig.~\ref{cmd71515}. The location in Fig.~\ref{cmd71515} of
these objects whose AGB nature is confirmed by IRAS fluxes supports
the consistency of the borderline between AGB stars and YSOs proposed
by Felli et al.(2000).  These ISO-IRAS objects are bright sources in the
ISOGAL catalogue since the ISOGAL sensitivity is much higher than
IRAS. However, because they have large A$_{\it V}$ and are either AGB stars with high mass loss
rate or YSOs embedded in cold circumstellar envelope, they also exhibit
very red colors in the DENIS bands. In spite of their brightness in LW2
and LW3 bands, no IRAS object was bright enough to be detectable in the I-band
by DENIS, except one, \#169, whose non-detection by DENIS was caused
by saturation other than by faintness.

\begin{table*}[ht]
\caption{ISOGAL sources cross-identified in the IRAS PSC catalogue.
These sources, except \#169, are too faint to be detected in the DENIS 
I-band. Their DENIS magnitudes, ISOGAL magnitudes and flux densities, IRAS
flux densities at 12, 25, 60 and 100$\mu$m, and MSX flux densities in band 
A(8$\mu$m), C(12$\mu$m), D(15$\mu$m) and
E(21$\mu$m) are listed in order.  }
\begin{tabular}{ccccccccccccccccc} \hline

ISO & IRAS~PSC & J & K$_{\rm S}$ & \multicolumn{2}{c}{[7]} &
 \multicolumn{2}{c}{[15]} & 12 & 25 & 60 & 100 & A & C & D & E &
 Type\\ & & mag & mag & mag & Jy & mag & Jy & Jy & Jy & Jy & Jy & Jy &
 Jy & Jy & Jy & \\ \hline 22 & 16459-4354 & -- & 10.9 & 4.4 & 1.4 &
 3.3 & 0.9& -- & 2.1 & -- & -- & 0.5 & -- & 1.2 & -- & AGB \\ 134 &
 16464-4359 & -- & -- & 5.3 & 0.6 & 3.9 & 0.5& -- & 9.0 &199.0 & -- &
 1.4 & -- & 1.0 & 3.3 & YSO\\ 169 & 16465-4344 &
 \multicolumn{2}{c}{Saturation}& 3.1 & 4.8& 2.8 & 1.4& 2.0 & -- & -- &
 -- & 2.4 & 1.6 & 1.2 & -- & ?\\ 204 & 16466-4353 & -- & 8.0 & 2.2 &
 11.5& 0.6 & 11.0& 4.3 & 6.3 & -- & -- & 6.4 & 7.6 & 8.2 & 7.1 & AGB\\
 324 & 16468-4349 & 15.5 & 8.9 & 3.9 & 2.3 & 2.6 & 1.8 & -- & 1.3 & --
 & -- & 0.6 & 1.1 & 1.2 & -- & AGB\\ 356 & 16469-4356 & 12.9 & 6.4 &
 3.2 & 4.4 & 1.1 & 7.2 & 5.2 & 4.7 & -- & -- & 3.0 & 6.0 & 5.8 & 4.0 &
 AGB\\ 364 & 16469-4348 & 12.0 & 7.7 & 6.0 & 0.3 & 4.0 & 0.5 & -- &
 1.6 & 64.7 &236.5 & 0.4 & -- & -- & 2.4 & YSO\\ 506 & 16472-4351 &
 -- & -- & 5.3 & 0.6 & 1.6 & 4.4 & 3.4 & 8.4 & -- & -- & -- & 2.0 &
 3.6 & 4.7 & YSO?\\ 565 & 16475-4349 & 14.3 & 7.0 & 2.3 & 10.0 & 1.3 &
 5.8 & 5.6 & 7.3 & -- & -- & 3.8 & 5.0 & 5.5 & 5.4 & AGB\\ \hline

\end{tabular}
\label{tabiras}
\end{table*}

The ISOGAL objects were also cross-identified with the MSX point
source catalogue available at http://www.ipac.caltech.edu/ipac/msx/msx.html.  The MSX (Midcourse Space eXperiment) surveyed the
entire Galactic plane between +4.5 \degr and --4.5\degr~ at 4.3$\mu$m(band B),
 8.3$\mu$m(band A),
12.1$\mu$m(band C), 14.6$\mu$m(band D) and 21.3{\rm $\mu$m} (band E)
(Egan et al. 1996; Price et al. 2001). The FC--01863+00035 ISOGAL field was in the MSX observing
area.  In order to cross identify the ISOGAL sources with the MSX
objects, the search radius was set to 10\arcsec ~which is about
4$\sigma$ position uncertainty of MSX (Cohen et al. 2000). The brightness
of the ISOGAL objects to be associated with MSX is also limited to
those with $[7]<7.0$ since the MSX SPIRIT III detector was much less
sensitive than ISOCAM, so that the possibility of spurious chance
associations is quite negligible. Twenty-five ISOGAL objects (4\%) are found
to have counterpart in the MSX PSC catalogue and they are all shown in
Fig.~\ref{cmd71515}. The MSX results of the objects associated with
IRAS PSC are listed in Table~\ref{tabiras}.  Most of them, with 
[15]$<$6.0 and 0.5$<$[7]--[15]$<$1.6, have MSX color indexes  
like luminous mass-losing AGB
stars, more or less distant. Three of them (\#306 (not in Table~\ref{tabiras}), \#364 and \#506, plus
\#134) are probably YSOs from their color [7]--[15]$>$1.8 and their
MSX colors ; they could be nearby or massive since they are all
brighter than 5 mag at [15]. 

A table with such data about all ISOGAL-MSX
associations is available at CDS (Table 5).
At least six blue sources with [7]--[15]$<$0.4 
(\#48, \#119, \#169, \#190, \#397 and \#587, see Fig.~\ref{cmd71515}) should be relatively nearby
stars with little circumstellar dust, either RGB, or early AGB, or
main sequence stars.

\section{Conclusion}

The combination of 7 and 15 $\mu$m ISOGAL data with near-infrared
DENIS data is powerful to trace the extinction and Galactic structures
on the lines of sight of ISOGAL fields in the inner Galactic disk. Our
detailed analysis of the good quality ISOGAL-DENIS data of the field
FC--01863+00035 has firstly allowed to derive a reasonably accurate value
of the extinction at 7 {\rm $\mu$m} through ${\rm A_{K_{S}}-A_7 = 0.35(A_J-A_{K_{S}})}$ which yields A$_{7}$/A$_{\it V}$ $\sim$0.03 from the
near-IR extinction values of van de Hulst--Glass (Glass 1999). One can thus use the
value of the color K$_{\rm S}$--[7] to determine the very large interstellar
extinction on the lines of sight of the ISOGAL sources without too much
circumstellar dust, in the range of A$_{\it V}$ $\sim$25--45 
where the DENIS data
alone fail because of non-detection in the J-band. The preliminary
comparison with CO and HI data confirms that most of the interstellar
extinction comes from the large molecular clouds in the Centaurus arm
which produces the bulk of the CO and HI emission. For sources with very
large extinction, future studies with higher-resolution CO data should
help to distinguish whether the excess of extinction mainly comes from a
condensation in this cloud or from additional dust layers farther out.

The combination of the ISOGAL and DENIS data and the knowledge of the
interstellar extinction allow an approximate determination of the
nature of most of the ISOGAL sources. As in other ISOGAL fields, the
bulk of ISOGAL sources are AGB stars with weak mass-loss and bright
RGB stars, with some high mass-loss AGB stars and of bluer and
less luminous foreground stars, together with a few YSOs. The
combination with MSX and IRAS data helps to identify more precisely
the brightest and reddest sources and to find the most interesting
ones for follow-up studies. The values of the colors (K$_{\rm S}$--[15])$_{0}$ and
[7]--[15] could provide reasonably accurate estimates of the mass-loss
rates of AGB stars (e.g. Ojha et al. 2002). The information about the
extinction structure on the line of sight may bring some information
on the distance of ISOGAL sources in the Galactic disk. But the
distances are still much too uncertain to allow accurate estimates of
the luminosities, and, hence, to elaborate, e.g., on the age of the
AGB populations. The main differences between this disk field and
Baade's Window are that there is a larger proportion of relatively
nearby objects, and that there is a branch of a few YSO
candidates. They either have a redder color index $[7]-[15]>1.6$ or
were detected only in the LW3 band. However, further observations 
are needed to confirm their nature.

There are obviously many possible  extensions of such a case
study of an ISOGAL field: systematic similar analysis of the
extinction in all other ISOGAL fields; extended studies of stellar populations in various disk fields
which is in progress (Felli et al. 2002); 
detailed combined studies with better CO data, and with continuum
mm/submm emission of molecular clouds; systematic analysis of
mass-loss of AGB stars of the inner Galactic disk; identification of
lists of particular interesting sources for spectroscopic follow-up studies, in
particular in the mid-IR with SIRTF, in the near-IR with various ground telescopes,  and in the visible for low--extinction sources; more sophisticated analysis with
multi-wavelength data at high angular resolution from the SIRTF/GLIMPSE 
legacy program and from 
wide-field near-IR cameras.

\acknowledgements{B.W. Jiang thanks NKBRSF G19990754 and the support
from CAS and CNRS, and Dr. Aigen~Li for useful discussion.  We particularly 
thank M.~Schultheis, F.~S\'evre and T.~August for
their help. We thank Dr. G.Gilmore for his helpful referee's comments and suggestions. S. Ganesh, G. Simon and A. Omont acknowledge support from Project N. 1910-1 of
the Indo--French collaboration CEFIPRA/IFCPAR. }

\bibliographystyle{aa}

\end{document}